\documentclass[review]{elsarticle}

\usepackage{lineno,hyperref}
\usepackage{physics}
\usepackage{amsmath,amssymb,amsfonts}
\usepackage{cleveref}
\usepackage{filecontents}
\usepackage{xcolor}
\usepackage{microtype}
\usepackage{nth}
\usepackage{mathtools}
\usepackage{paralist}
\usepackage{pgfplots}
\usepackage{tikz}
\modulolinenumbers[5]

\journal{Computer Physics Communications}

\usepackage{indentfirst} 
\usepackage[utf8]{inputenc}
\usepackage{graphicx}
\usepackage{algorithmic} 
\usepackage[caption=false,font=footnotesize]{subfig}
\usepackage[export]{adjustbox}
\usepackage{siunitx}

\title{Acceleration techniques for semiclassical Maxwell-Bloch systems: An application to discrete quantum dot ensembles}

\definecolor{msugreen}{RGB}{24,69,59}

\newcommand{\naive}{na\"{\i}ve}

\newcommand{\ie}{\emph{i.e.}}     

\crefname{appsec}{appendix}{appendices}

\newcommand{\integralspace}{\kern0.1em}
\newcommand{\sint}{\int_{} \integralspace}

\DeclarePairedDelimiter\floor{\lfloor}{\rfloor}

\newcommand{\attrib}{} 

\definecolor{cbred}{HTML}{e41a1c}
\definecolor{cbblue}{HTML}{377eb8}
\definecolor{cbgreen}{HTML}{4daf4a}

\newif\ifmakeplots
\makeplotstrue

\newcommand{\qd}{quantum dot}
\newcommand{\qds}{quantum dots}

\newcommand{\outerprod}[2]{#1 \! \otimes \! #2}
\newcommand{\tensor}[1]{#1}

\pgfplotsset{compat=1.4}

\usepgfplotslibrary{colorbrewer}
\pgfplotsset{cycle list/Set1-5}

\begin{document}
\begin{frontmatter}


\author{C. Glosser, E. Lu, T. J. Bertus, C. Piermarocchi and B. Shanker}
\address{Department of Physics \& Astronomy, Michigan State University 567 Wilson Road, East Lansing, Michigan 48824, USA\\
Department of Electrical \& Computer Engineering, Michigan State University 428 South Shaw Lane, East Lansing, Michigan 48824, USA}




\begin{abstract}
The solution to Maxwell-Bloch systems using an integral-equation-based framework has proven effective at capturing collective features of laser-driven and radiation-coupled \qds, such as light localization and modifications of Rabi oscillations~\cite{glosser2017collective}.
Importantly, it enables observation of the dynamics of each \qd\ in large ensembles in a rigorous, error-controlled, and self-consistent way \emph{without} resorting to spatial averaging.
Indeed, this approach has demonstrated convergence in ensembles containing up to $10^4$ interacting \qds~\cite{glosser2017collective}.
Scaling beyond $10^4$ \qds\ tests the limit of computational horsepower, however, due to the $\mathcal{O}(N_t N_s^2)$ scaling (where $N_t$ and $N_s$ denote the number of temporal and spatial degrees of freedom).
In this work, we present an algorithm that reduces the cost of analysis to $\mathcal{O}(N_t N_s \log^2 N_s)$.
While the foundations of this approach rely on well-known particle-particle/particle-mesh and adaptive integral methods, we add refinements specific to transient systems and systems with multiple spatial and temporal derivatives.
Accordingly, we offer numerical results that validate the accuracy, effectiveness and utility of this approach in analyzing the dynamics of large ensembles of \qds.
\end{abstract}


\end{frontmatter}

\maketitle

\section{Introduction\label{sec:Intro}}

The computational simulation of the nonlinear propagation of laser pulses through materials presents formidable challenges, particularly in materials containing dispersed \qds\ or nanoparticles that have strong light-matter coupling. One phenomenon, Rabi oscillations, demonstrates nonlinear behavior that can arise from such coupling.
These oscillations have a long history of study in single \qds{}~\cite{Stievater2001,Kamada2001,Htoon2002}, though understanding the collective Rabi dynamics of \qd\ \emph{ensembles} requires a careful analysis of emission effects that couple \qds\ to produce many-body collective effects.
Researchers have a significant interest in examining these effects from theoretical, numerical, and experimental perspectives~\cite{Slepyan2002,Slepyan2004,glosser2017collective,Asakura2013}. For instance, experiments on novel systems based on perovskite nanocrystals have recently demonstrated these effects~\cite{raino2018superfluorescence}. These experiments could lead to novel composite materials with enhanced optical properties. 

Typical theoretical and computational analyses use variations of the Maxwell-Bloch equations~\cite{Gross1982} to describe the collective behavior of ensembles of optically active centers in which classical radiation fields couple a quantum description of each center. To this end, methods such as homogenization~\cite{McCall1969,Rehler1971}, differential-equation-based methods \cite{Bachelard2015,Fratalocchi2008,Vanneste2001,baczewskiThesis,sytnyk2009}, and, more recently, integral-equation-based methods~\cite{glosser2017collective} all describe the dynamics of coupled \qd\ systems, each with a differing level of fidelity.

Approaches that do not rely on homogenization use coupled discrete methods to solve the classical Maxwell equations and local time evolution techniques to solve the Bloch equations for each \qd.
Spatial homogenization, on the other hand, describes the near and far radiation characteristics of a \qd\ assuming homogeneous background material properties~\cite{Temnov2009}.
This approach has limited validity as it does not account for strong interactions between particles in each other's nearfield, a shortcoming exacerbated by the non-linear regimes considered here.
Differential equation methods~\cite{Jin2002} to solve the Maxwell system have included finite-difference, time-domain finite-element, and discontinuous Galerkin methods, though all succumb to various numerical inaccuracies due to the nature of discretization. The inaccuracies most pertinent to simulation of quantum dot systems extend from the need to include point dipole sources in the simulation and capture near field effects that behave as $1/r^3$ (where $r$ denotes the distance between centers). Accurately recovering these fields has numerous challenges and one needs dense discretization in the vicinity of dots together with equivalent/soft sources to accurately capture these effects \cite{baczewskiThesis}. Additionally, static null spaces that grow linearly with time present another challenge with conventional time-domain finite-element techniques \cite{Jin2002}. On the plus side, these methods offer a high degree of flexibility and can accommodate different background linear bulk materials.

Our approach~\cite{glosser2017collective} differs significantly.
We make use of an integral equation-based formulation that employs a retarded potential to compute fields radiated by every \qd.
This approach does not rely on a particular discretization and only depends on the number of \qds\ under investigation.
Demonstrations of the accuracy of this approach appear in~\cite{glosser2017collective}, though this method faces a fundamental bottleneck: both the memory necessary and computational cost scale as $\mathcal{O}(N_t N_s^2)$ which becomes prohibitive for extended ensembles.
A traditional acceleration technique that readily adapts to these equations makes use of a rotated frame approximation to reduce the number of timesteps by a factor of $\sim\!1000$.
This approximation exploits the narrowband nature of the nonlinearity which enables the use of envelope functions \cite{sytnyk2009}. Physically, this arises due to the large difference between the characteristic Rabi energy associated with light-matter coupling and the optical transition energy. Computationally, this enables timestep sizes much larger than the inverse of the laser frequency. While this approximation offers significant acceleration, the key bottleneck \uppercase{\textbf{is}} the cost of evaluating retarded potentials, which scales \emph{quadratically} with the number of \qds; thankfully, there exists extensive literature on reducing this cost that we examine next. 

Two algorithms typically see use in accelerating the evaluation of retarded potentials: the Plane-Wave Time-Domain method (PWTD) and the Adaptive Integral Method (AIM, a variation of particle-particle/particle-mesh techniques) \cite{shanker2003fast,Yilmaz2004}. Unfortunately, neither of these methods apply directly to the problem at hand. To set the stage for discussion, assume that one needs to evaluate the radiated electric field $\mathbf{E}_R(\mathbf{r},t)= \mathfrak{F}\{\vb{P}(\vb{r}, t)\} $ \cite{Kobidze2005} due to a polarization density $\mathbf{P} (\mathbf{r},t)$ via $\mathfrak{F}\{\vb{P}(\vb{r}, t)\} =  \mathcal{L}\qty{g(\vb{r},t)} \star_{st} \mathbf{P}(\mathbf{r},t)$ where $g(\vb{r},t) = \delta(t - |\vb{r}|/c)/(4 \pi|\vb{r}|)$ denotes the retarded potential, $\mathcal{L}\qty{g(\vb{r},t)} = -\mu_0 \qty (\partial_t^2 \tensor{\mathrm{I}}\cdot - c^2 \nabla \nabla \cdot)g(\vb{r},t)$ denotes a dyadic differential operator, and $\tensor{\mathrm{I}}$ denotes the identity dyad. 

PWTD exploits the properties of radiated fields due to quiescent sources that occupy a bounded spatial domain. These fields have a bandlimit (in momentum space) that PWTD leverages to reconstruct them to arbitrary precision using a tree-based approach; see \cite{shanker2003fast} and references therein. However, the crux of this methodology lies in the fact that spatial variation scales with the temporal one (times $c$). Unfortunately, while this holds in the fixed frame, it does not in the rotated frame. 

AIM, on the other hand, relies on moments around a uniform grid (independent of temporal variation) to reconstruct sources and their resulting field distributions. This idea---using grids for computing translationally-invariant functions by exploiting an underlying block Toeplitz structure---has seen extensive use in molecular dynamics simulations to evaluate Laplace, Helmholtz, and wave equation kernels \cite{Bleszynski1996,Yilmaz2004,Rapaport2004}. In all these cases, one evaluates the space (and time) convolution with a scalar quantity, sans the operator $\mathcal{L}\qty {\cdot}$ required herein. Unfortunately, the nature of $\mathcal{L}\qty{\cdot}$ determines the number of potentials that need evaluation. As we will see in the ensuing sections, the $\mathcal{L}\qty{\cdot}$ used for quantum dots contains numerous dyadic terms, each with different orders of temporal derivatives. As a result, a \naive\ application of AIM to each term of the expression quickly becomes untenable and we seek to develop a more expedient technique.

This paper has two principal contributions:
\begin{inparaenum}[(i)]
    \item development and demonstration of techniques that overcome computational complexity (memory and CPU costs) associated with evaluation of integral equation operators, and 
    \item demonstration of these algorithms to examine optical systems containing \qds.
\end{inparaenum}
In developing these algorithms, we examine their runtime and accuracy and, more importantly, show that evaluating $\mathcal{L}\qty{g(\vb{r},t)}$ incurs approximately the same cost as evaluating $g(\vb{r},t)$ albeit with lower error.

We organize the rest of this paper as follows: in \cref{sec:problemStatement} we define the problem, and provide the means to a solution in \cref{sec:classical}.
\Cref{sec:AIM} develops the AIM method for the Maxwell-Bloch problem and outlines its computational complexity.
Next, in \cref{sec:results}, we present a number of results that verify the claims of accuracy, complexity, and applicability of this method to a collection of \qds.
Finally, in \cref{sec:conclusions} we summarize the contributions of the paper and outline future research avenues.

\section{Formulation\label{sec:problemStatement}}
Consider a domain $\Omega$ that contains $N_s$ randomly distributed \qds{}.
A time-varying electromagnetic field of central frequency $\omega$ impinges on $\Omega$ and excites each \qd{}.
We wish to develop the means to study the evolution of these \qds{} in response to both the incident excitation as well as radiation produced by other \qds{}.
Toward this end, we employ  a semi-classical approach to understand the response of each \qd{} to the incident field that comprises the laser field as well as fields radiated by other dots (computed classically).
In what follows, we provide a brief description of the requisite formulation for completeness; \cite{glosser2017collective} provides a more detailed description.

Dipolar transitions govern the response of each \qd{} to the exciting field. Specifically, we write the time-dependence of a given \qd's density matrix, $\hat{\rho}(t)$, as
\begin{equation}
  \dv{\hat{\rho}}{t} = \frac{-i}{\hbar}\commutator{\hat{\mathcal{H}}(t)}{\hat{\rho}} - \hat{\mathcal{D}}\qty[\hat{\rho}].
  \label{eq:liouville}
\end{equation}
For two-level systems, $\hat{\rho}(t)$ denotes a two by two matrix with three unique unknowns ($\rho_{00}$ and the real and imaginary parts of $\rho_{01}$), $\hat{\mathcal{H}}(t)$ represents a local Hamiltonian that governs the internal two-level structure of the \qd{} as well as its interaction with an external electromagnetic field, and $\hat{\mathcal{D}}$ provides dissipation terms that account for spontaneous emission effects phenomenologically.
Explicitly,
\begin{subequations}
  \begin{align}
    \hat{\mathcal{H}}(t) &\equiv \mqty(0 & \hbar \chi(t) \\ \hbar \chi^*(t) & \hbar \omega_0) \label{eq:hamiltonian}\\
    \hat{\mathcal{D}}\qty[\hat{\rho}] &\equiv \mqty( \qty(\rho_{00} - 1)/{T_1} & \rho_{01}/{T_2} \\ \rho_{10}/{T_2} & \rho_{11}/T_1 ) \label{eq:dissipator}
  \end{align}
  \label{eq:qd operators}
\end{subequations}
where $\chi(t) \equiv \vb{d} \cdot \hat{\vb{E}}(\vb{r}, t)/\hbar$, $\vb{d} \equiv \matrixel{1}{e \hat{\vb{r}}}{0}$, and the kets represent the highest valence and lowest conduction states of the \qd{} under consideration.
Finally, the $T_1$ and $T_2$ constants characterize average relaxation  and decoherence times.

We compute the semi-classical interaction between \qds{} assuming coherent fields and negligible quantum statistical effects.
Such assumptions imply classical electromagnetic interactions while preserving the two-level structure of individual \qds{}.
To this end, we write the total electric field at any point in space and time as $\vb{E}(\vb{r}, t) = \vb{E}_L(\vb{r}, t) + \mathfrak{F}\{ \vb{P}(\vb{r}, t) \}$
where $\vb{E}_L(\vb{r}, t)$ denotes the incident laser field, $\vb{P}(\vb{r}, t)$ gives a polarization distribution arising from the off-diagonal elements (coherences) of $\hat{\rho}$, and
\begin{equation}
\begin{split}
    \mathfrak{F}\{\vb{P}(\vb{r}, t)\} & \doteq - \mu_0 \qty (\partial^2_t \tensor{\mathrm{I}} - c^2 \nabla \nabla ) g \qty (\mathbf{r},t) \star_{st} \mathbf{P} \qty (\mathbf{r},t ) \\
    & \doteq \frac{-1}{4\pi \epsilon} \int_{}
      \left(\tensor{\mathrm{I}} -  \outerprod{\bar{\vb{r}}}{\bar{\vb{r}}} \right) \cdot \frac{\partial_t^2 \vb{P}(\vb{r}', t_R)}{c^2 \abs{\vb{r}-\vb{r}'}} +
      \left(\tensor{\mathrm{I}} - 3\outerprod{\bar{\vb{r}}}{\bar{\vb{r}}} \right) \cdot \qty(
        \frac{\partial_t   \vb{P}(\vb{r}', t_R)}{c \abs{\vb{r}-\vb{r}'}^2} +
        \frac{             \vb{P}(\vb{r}', t_R)}{  \abs{\vb{r}-\vb{r}'}^3}
      ) \dd[3]{\vb{r}'}
      \end{split}
  \label{eq:integral operator}
\end{equation}
(see~\cite[section \S 72]{Landau2013}).
In the above expression,  $\bar{\vb{r}} \equiv \qty(\vb{r} - \vb{r}')/\abs{\vb{r} - \vb{r}'}$, $\otimes$ represents the tensor product (i.e.\ $\qty(\outerprod{\vb{a}}{\vb{b}})_{ij} = a_i b_j$), $t_R \equiv t - \abs{\vb{r} - \vb{r}'}/c$, and $\epsilon$ gives the dielectric constant of the inter-dot medium.
Thus, in a system composed of multiple \qds{}, \cref{eq:integral operator} couples the evolution of each \qd{} by way of the off-diagonal matrix elements appearing in \cref{eq:hamiltonian}.

Note that this approach does not require an instantaneous dipole-dipole Coulomb term between (charge-neutral) \qds{}; the interactions between structures occur only via the electric field which propagates through space with finite velocity (see ~\cite[sections {A}$_{\textsc{iv}}$ and {C}$_{\textsc{iv}}$]{Cohen1989} for in-depth discussions of this point).

In the systems under consideration here, $\omega_0$ lies in the optical frequency band ($\sim \SI{1500}{\milli\eV\per\hbar}$).
Consequently, integrating \cref{eq:liouville} directly to resolve the Rabi dynamics that occur on the order of \SI{1}{\pico\second} quickly becomes computationally infeasible.
Introducing $\tilde{\rho} = \hat{U} \hat{\rho} \hat{U}^\dagger$ where $\hat{U} = \mathrm{diag}(1, e^{i \omega t})$, we may instead write \cref{eq:liouville} as
\begin{equation}
  \dv{\tilde{\rho}}{t} = \frac{-i}{\hbar} \commutator{\hat{U} \hat{\mathcal{H}} \hat{U}^\dagger - i \hbar \hat{V}}{\tilde{\rho}} - \hat{\mathcal{D}}\qty[\tilde{\rho}], \quad \hat{V} \equiv \hat{U} \dv{\hat{U}^\dagger}{t}
  \label{eq:rotating liouville}
\end{equation}
which contains only terms proportional to $e^{i (\omega_0 \pm \omega) t}$ if $\vb{E}(t) \sim \tilde{\vb{E}}(t)\cos(\omega t)$.
Consequently, we ignore the high-frequency quantities (corresponding to $\omega_0 + \omega$) under the assumption that such terms will integrate to zero in solving \cref{eq:rotating liouville} over appreciable timescales~\cite{Allen1975}. 
One can then construct efficient numerical strategies for solving \cref{eq:rotating liouville}.
A similar transformation applies to the source distribution $\vb{P}(\vb{r}, t)$; by assuming $\vb{P}(\vb{r}, t) = \tilde{\vb{P}}(\vb{r}, t)e^{i \omega t}$ in \cref{eq:integral operator} the radiated field envelope becomes
\begin{equation}
\label{eq:radiated envelope}
  \begin{gathered}
    \tilde{\mathfrak{F}}\{ \tilde{\vb{P}}(\vb{r}, t) \} \equiv \frac{-1}{4\pi \epsilon} \int_{}
    \qty(\tensor{\mathrm{I}} -  \outerprod{\bar{\vb{r}}}{\bar{\vb{r}}} ) \cdot \frac{\qty(\partial_t^2 \tilde{\vb{P}}(\vb{r}', t_R) + 2 i \omega \partial_t \tilde{\vb{P}}(\vb{r}', t_R) - \omega^2 \tilde{\vb{P}}(\vb{r}', t_R)) e^{-i \omega \abs{\vb{r} - \vb{r}'}/c}}{c^2 \abs{\vb{r}-\vb{r}'}} + \\
    \qty(\tensor{\mathrm{I}} - 3\outerprod{\bar{\vb{r}}}{\bar{\vb{r}}} ) \cdot \frac{\qty(\partial_t \tilde{\vb{P}}(\vb{r}', t_R) + i \omega \tilde{\vb{P}}(\vb{r}', t_R))e^{-i \omega \abs{\vb{r} - \vb{r}'}/c}}{c \abs{\vb{r}-\vb{r}'}^2} +
    \qty(\tensor{\mathrm{I}} - 3\outerprod{\bar{\vb{r}}}{\bar{\vb{r}}} ) \cdot \frac{                \tilde{\vb{P}}(\vb{r}', t_R) e^{-i \omega \abs{\vb{r} - \vb{r}'}/c}}{\abs{\vb{r}-\vb{r}'}^3}
  \, \dd[3]{\vb{r}'}.
  \end{gathered}
\end{equation}
Note that \cref{eq:radiated envelope}  maintains the high-frequency phase relationship between sources oscillating at $\omega$ via the factors of $e^{-i \omega \abs{\vb{r} - \vb{r}'}/c}$ that appear.
As a result, we write
\begin{equation}\label{eq:rotatedRadFld}
    \tilde{\vb{E}}(\vb{r}, t) = \tilde{\vb{E}}_L(\vb{r}, t) + \tilde{\mathfrak{F}}\{ \tilde{\vb{P}} (\vb{r}, t) \}
\end{equation}
and the evolution of the ensemble relies on a self-consistent solution of \cref{eq:rotating liouville,eq:rotatedRadFld}. As evident from \cref{eq:radiated envelope}, this comprises a large number of costly potential integrals arising from the number of dyadic components and number of time derivatives. As such, we now turn our attention to an efficient computational infrastructure for ameliorating this cost.
\section{Discrete Solution}
\label{sec:classical}

Solving \cref{eq:rotating liouville,eq:rotatedRadFld} self-consistently proceeds via the following steps:
\begin{inparaenum}[(i)]
  \item represent the time varying behavior of the polarization
  \item using \eqref{eq:rotatedRadFld}, evaluate $\tilde{\vb{E}}(\vb{r}, t)$ at a given timestep, and
  \item use a predictor corrector approach to evaluate $\tilde{\rho}$ via \eqref{eq:rotating liouville}.
\end{inparaenum}
Representing $\tilde{\vb{P}}(\vb{r}, t)$ in terms of space and time basis functions such that
\begin{equation}
  \tilde{\vb{P}}(\vb{r}, t) \approx \sum_{\ell = 0}^{N_s-1} \sum_{m = 0}^{N_t - 1} \tilde{\mathcal{A}}_\ell^{(m)} \vb{s}_\ell(\vb{r}) T(t - m \, \Delta t),
  \label{eq:discretization}
\end{equation}
$\tilde{\mathcal{A}}_\ell^{(m )} = \tilde{\rho}_{\ell,01}(m \, \Delta t)$
gives the polarization associated with the $\ell$th \qd{} at the $m$th time step, and $\Delta t$ denotes a fixed time interval chosen to accurately sample the dynamics of the physical quantities involved. Both $\vb{s}_\ell(\vb{r})$ and $T(t)$ have finite support and  $T(t)$ obeys (discrete) causality (i.e.\ $T(t) = 0$ if $t < -\Delta t$).
In particular, we consider shifted Lagrange polynomials for the $T(t)$ and assume dipolar transitions in the \qds{} allowing for $\vb{s}_\ell(\vb{r}) = \vb{d}_\ell \delta(\vb{r} - \vb{r}_\ell)$, though this analysis readily extends to accommodate any similar set of functions \cite{Pray2012,Shanker2000}.

Substituting \cref{eq:discretization} into \cref{eq:radiated envelope} and projecting the resulting fields onto $\delta(t - m \, \Delta t) \vb{s}_\ell(\vb{r})$, we obtain
\begin{equation}
  \tilde{\mathcal{E}}^{(m)} = \tilde{\mathcal{E}}_\text{inc}^{(m)} + \sum_{m'= 0}^{m} \tilde{\mathcal{F}}^{(m - m')} \boldsymbol{\cdot} \tilde{\mathcal{A}}^{(m')}
  \label{eq:mot}
\end{equation}
where
\begin{subequations}
  \begin{align}
    \tilde{\mathcal{E}}^{(m)} &= \langle \vb{s}_\ell(\vb{r}), \, \tilde{\vb{E}}(\vb{r}, m \, \Delta t) \rangle \label{eq:f elements}\\
    \tilde{\mathcal{E}}^{(m)}_{\text{inc},\ell} &= \langle \vb{s}_\ell(\vb{r}), \, \tilde{\vb{E}}_\text{inc}(\vb{r}, m \, \Delta t) \rangle \label{eq:l elements}\\
    \tilde{\mathcal{F}}^{(k)}_{\ell\ell'} &= \big\langle \vb{s}_\ell(\vb{r}), \, \tilde{\mathfrak{F}}\qty{\vb{s}_{\ell'}(\vb{r}) T(k \, \Delta t)} \big\rangle \label{eq:z elements}
  \end{align}
\end{subequations}
and $\langle \cdot, \cdot \rangle$ denotes the inner product between functions.

A self-consistent solution to \cref{eq:rotating liouville,eq:rotatedRadFld} then has the following prescription for any timestep:
\begin{inparaenum}[(i)]
  \item predict  $\tilde{\mathcal{A}}_\ell^{(m )} = \tilde{\rho}_{\ell,01}(m \, \Delta t)$ $\forall~\ell$ from the known prior history of the system, 
  \item compute $\tilde{\mathcal{E}}^{(m )}_\ell$ using \cref{eq:mot},
  \item find $\partial_t \, \tilde{\rho}_{\ell,01} (m \, \Delta t)$ using \cref{eq:rotating liouville}, and
  \item correct $\tilde{\rho}_{\ell,01} (m \, \Delta t)$ and iterate through steps (ii) through (iv) until converged.
\end{inparaenum}

The time complexity of the entire algorithm follows naturally from the above description: for $N_s$ particles and $N_t$ timesteps, the cost of evaluating \cref{eq:mot} scales as $\mathcal{O}(N_t N_s^2)$ while the cost of solving \cref{eq:rotating liouville} for every \qd{} scales as $\mathcal{O}(N_t N_s)$.
As a result, the bottleneck arises from the discrete convolution/field evaluation at every timestep, and we address strategies to ameliorate this cost in the next section.
\section{Acceleration via Fast Fourier Transforms\label{sec:AIM}}

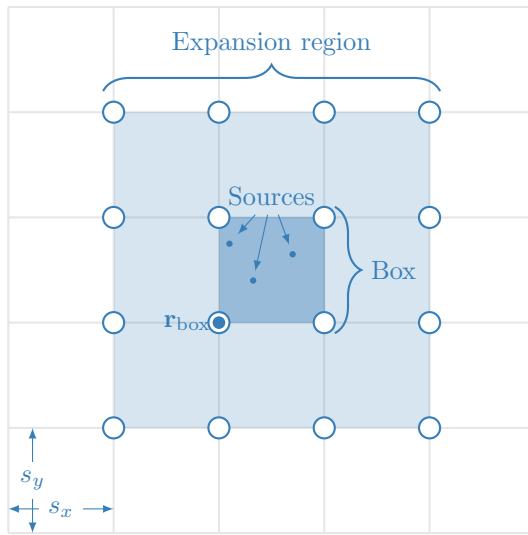
\begin{figure}
  \centering
  \usetikzlibrary{colorbrewer, decorations.pathreplacing, shapes, calc}

\begin{tikzpicture}[>=latex, scale=1.4]

\draw[step=1, black, thick, opacity=0.1] (-2,-2) grid (3, 3);

\fill[cbblue, opacity=0.4] (0,0) rectangle (1, 1);
\fill[cbblue, opacity=0.2] (-1,-1) rectangle (2, 2);

\foreach \x in {-1,...,2} {
  \foreach \y in {-1,...,2} {
    \draw[thick, cbblue, fill=white, radius=0.1] (\x,\y) circle;
  }
}



\node[radius=0.03] (g1) at (0.25, 0.25) {};
\node[radius=0.03] (g2) at (0.50, 0.25) {};
\node[radius=0.03] (g3) at (0.75, 0.25) {};



\fill[cbblue, radius=0.03] (0.1, 0.75) node (A) {} circle;
\fill[cbblue, radius=0.03] (0.325, 0.4) node (B) {} circle;
\fill[cbblue, radius=0.03] (0.7, 0.65) node (C) {} circle;

\fill[cbblue, radius=0.06] (0, 0) circle;
\node[cbblue] at (-0.3, 0) {$\mathbf{r}_\text{box}$};

\node[cbblue] (pts) at (0.5, 1.2) {Sources};
\draw[cbblue, ->] (pts) -- (A);
\draw[cbblue, ->] (pts) -- (B);
\draw[cbblue, ->] (pts) -- (C);

\draw[cbblue, thick, decorate,decoration={brace,amplitude=10}] (-1.1,2.2) -- (2.1, 2.2) node [midway, above, yshift=10] {Expansion region};
\draw[cbblue, thick, decorate,decoration={brace,amplitude=10}] (1.1,1.1) -- (1.1, -0.1) node [midway, right, xshift=10] {Box};

\draw[<->, cbblue] (-2, -1.77) -- (-1, -1.77) node [midway, fill=white] {$s_x$};
\draw[<->, cbblue] (-1.77, -2) -- (-1.77, -1) node [midway, fill=white] {$s_y$};

\end{tikzpicture}
  \caption{\label{fig:aim terminology} Illustration of the grid structure and related terminology.
    All of the sources within a box (shown as the central shaded square) map to the same set of expansion points (shown as open circles) indexed relative to $\vb{r}_\text{box}$.
  }
\end{figure}

As alluded to earlier, TD-AIM forms the basis our approach to reducing the computational complexity. Unfortunately, we cannot directly apply existing methodologies due to overhead induced by the multiplicity of terms as well as temporal derivatives. In what follows, we develop a variation of TD-AIM that relies on propagating the convolution of the retarded potential with the source function and \emph{local} evaluation of spatial and temporal derivatives.

\subsection{Algorithmic Details}

\begin{figure}
    \centering
    \usetikzlibrary{patterns}

\begin{tikzpicture}[>=latex]

\draw[step=1, black, thick, opacity=0.1] (-4,-2) grid (7, 3);

\fill[pattern=north east lines, pattern color=cbblue, opacity=0.2] (-3,-1) rectangle (0, 2);
\fill[cbblue!40!white] (-2,0) rectangle (-1, 1);

\foreach \x in {-3,...,0} {
  \foreach \y in {-1,...,2} {

    \draw[cbblue, fill=white, draw opacity=0.4, radius=0.1] (\x,\y) circle;

  }
}
\fill[cbblue, radius=0.06] (-2, 0) node[anchor=north east] () {} circle;

\fill[pattern=north west lines, pattern color=cbblue, opacity=0.2] (-1,-1) rectangle (2, 2);
\fill[cbblue!40!white] (0,0) rectangle (1, 1);

\foreach \x in {-1,...,2} {
  \foreach \y in {-1,...,2} {

    \draw[cbblue, fill=white, draw opacity=0.4, radius=0.1] (\x,\y) circle;

  }
}
\fill[cbblue, radius=0.06] (0, 0) node[anchor=north east] () {} circle;

\fill[pattern=north west lines, pattern color=cbblue, opacity=0.2] (3,-1) rectangle (6, 2);
\fill[cbblue!40!white] (4,0) rectangle (5, 1);

\foreach \x in {3,...,6} {
  \foreach \y in {-1,...,2} {

    \draw[cbblue, fill=white, draw opacity=0.4, radius=0.1] (\x,\y) circle;

  }
}
\fill[cbblue, radius=0.06] (4, 0) node[anchor=north east] () {} circle;

\fill[cbblue] (0.3, 0.7) node (A) {} circle (0.05);
\fill[cbblue] (-1.2, 0.2) node (B) {} circle (0.05);
\fill[cbblue] (4.55, 0.525) node (dest) {} circle (0.05);

\draw[thick, ->, cbred, opacity=0.8] (A) -- (0, 0.97) -- (4, 0.97) -- (4.5, 0.47);
\draw[thick, ->, cbblue, dashed] (A) -- (dest);
\draw[thick, ->, cbblue] (B) -- (0, 1.03) -- (4, 1.03) -- (4.5, 0.53);

\draw[cbblue, thick, decorate,decoration={brace,amplitude=10}] (-3.1,2.1) -- (0.1, 2.1) node [midway, above, yshift=8] {Box $A$};
\draw[cbblue, thick, decorate,decoration={brace,amplitude=10}] (2.1,-1.1) -- (-1.1, -1.1) node [midway, below, yshift=-4.5] {Box $B$};
\draw[cbblue, thick, decorate,decoration={brace,amplitude=10}] (6.1,-1.1) -- (2.9, -1.1) node [midway, below, yshift=-4.5] {Box $C$};

\end{tikzpicture}
    \caption{\label{fig:nearfield correction}Illustration of nearfield corrections between close boxes.
    Expansions in boxes $A$ and $B$ overlap, but only box $B$ lies in the nearfield of box $C$ for $\gamma=2$.
    The grid-based propagation strategy only remains accurate for distant source/observer pairs.
    To avoid incurring undue error, we remove the interaction ``through the grid'' between the $BC$ pair (red line) and replace it with a more accurate ``direct'' interaction (dashed blue line).
    The $AC$ pair requires no such treatment as they have well-separated expansion regions.}
\end{figure}
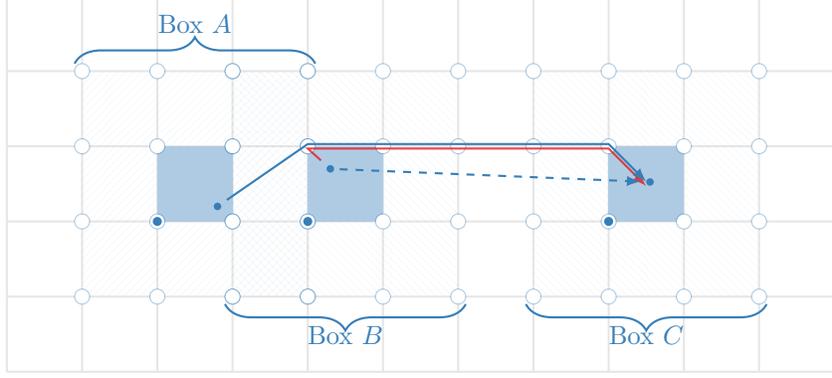

In what follows, we give algorithmic steps for the fixed frame elements, $\mathcal{F}^{(k)}$. The algorithmic steps for the rotating wave elements, $\tilde{\mathcal{F}}^{(k)}$, proceed identically. To effect a sub-quadratic calculation of \cref{eq:mot}, we approximate $\mathcal{F}^{(k)}$ as a sum of near- and far-field contributions.   The near-field matrix elements follow directly from \cref{eq:z elements}---sources within a prescribed distance threshold interact ``directly'' so as to avoid incurring unreasonable approximation error between adjacent basis functions. Sources beyond this threshold, however, interact via auxiliary spatial basis functions that reside at the vertices of a regular Cartesian grid. These auxiliary sources recover $\mathfrak{F}\qty{\vb{P}(\vb{r}, t)}= \mathcal{L}\qty { g(\vb{r}, t)} \star_{st} \vb{P}(\vb{r}, t)$ at large distances and have two computational advantages: (i) they compress the interaction matrix by representing sources within the same spatial region in terms of the same auxiliary set (\cref{fig:aim terminology}) and (ii) they impose a Toeplitz structure on the resulting interaction matrix that lends itself to efficient diagonalization through application of an FFT. Mathematically, 
\begin{equation}
  \begin{aligned}
    \mathcal{F}^\qty(m - m') & \approx \mathcal{F}^\qty(m-m')_\text{direct} + \Lambda_\mathfrak{F} \mqty(\partial_t^0 \mathcal{G}^\qty(m-m') \\ \partial_t^1 \mathcal{G}^\qty(m-m') \\ \partial_t^2 \mathcal{G}^\qty(m-m') \\ \vdots) \Lambda^\dagger \\
                             & \equiv \mathcal{F}^\qty(m-m')_\text{direct} + \mathcal{F}^\qty(m-m')_\text{FFT}
    \end{aligned}
  \label{eq:f decomposition}
\end{equation}
where
\begin{subequations}
  \begin{align}
    \mathcal{F}^\qty(m-m')_{\text{direct},\ell\ell'} &= \begin{cases}
      \mathcal{F}^\qty(m-m')_{\ell\ell'}  - \mathcal{F}^\qty(m-m',\tau)_{\text{FFT},\ell\ell} & R_{\ell\ell'} \leqslant \gamma \\
      0 & \text{otherwise},
    \end{cases} \\
    \mathcal{G}^\qty(m-m')_{ab} &= \left< \vb{u}_a(\vb{r})\delta\qty\big(t - (m-m')\, \Delta t), g(\vb{r}, t) \ast \vb{u}_{b}(\vb{r})T(t) \right>
  \end{align}
  \label{eq:decomposition elements}
\end{subequations}
The $\Lambda$ matrices in \cref{eq:f decomposition} denote the (sparse) projections to and from the grid (detailed in \cref{sec:expansion matrices}), and $\vb{u}_a(\vb{r})$ indicates an auxiliary basis function on the spatial grid indexed by $a$. Finally, $\tau_\text{max}$ and $\gamma$ serve as adjustable input parameters to control the accuracy of the simulation and $R_{\ell\ell'}$ gives the minimum distance (in integral units of the grid spacing) between the expansion regions enclosing $\vb{s}_\ell(\vb{r})$ and $\vb{s}_{\ell'}(\vb{r})$ (\cref{fig:nearfield criterion}) via
\begin{equation}
  R^\text{grid}_{\ell\ell'} = \min\!\qty{\norm{\vb{u} - \vb{u}'}_\infty \, | \, \vb{u} \in C_\ell, \, \vb{u}' \in C_{\ell'}}.
\end{equation}
Computationally, at every time step, our algorithm proceeds as follows: 
\begin{enumerate}
  \item \underline{Projection on to the uniform grid}: At timestep $m$ project each of the $\vb{s}_\ell(\vb{r}) \tilde{\mathcal{A}}^\qty(m)_\ell$ onto the auxiliary sources.
    Aside from discretization/sampling criteria, the operators in \cref{eq:mot} do not affect these projections, thus the distribution of auxiliary sources on the grid, that we indicate as $\tilde{\vb{P}}_\text{aux}(\vb{r}, t)$, mimics the distribution of $\tilde{\vb{P}}(\vb{r}, t)$ at large distances.
  \item \underline{Effect the convolution in \cref{eq:mot} between auxiliary sources}: 
    Having imposed a regular structure on $\tilde{\vb{P}}_\text{aux}(\vb{r}, t)$, we may efficiently diagonalize the matrix representing this (discrete) convolution with (up to four-dimensional) blocked FFTs.
    Note that the algorithm thus far has essentially evaluated the potential, $g(\vb{r}, t) \ast \tilde{\vb{P}}(\vb{r}, t)$, at $t = m \, \Delta t$ at every point $\vb{u}$ in the grid.
  \item \underline{Projection back from the grid}: Recover the total field under the action of $\mathfrak{F}$ by projecting the potential on each $\vb{u}$ back onto the $\vb{s}_\ell(\vb{r})$.
    These projections make use of specialized projection matrices that depend on the  derivatives contained inside $\mathfrak{F}$.
  \item \underline{Correction of near fields}: Subtract the fields determined by steps 1-3 for pairs of spatial basis functions within a prescribed distance threshold and replace it with \cref{eq:z elements}.
    The auxiliary grid approximations only remain accurate at large distances, thus this step corrects large approximation errors that occur between adjacent $\vb{s}_\ell(\vb{r})$. (\Cref{fig:nearfield correction} gives a schematic illustration of this correction.)
\end{enumerate}

\begin{figure}
  \centering
  \usetikzlibrary{colorbrewer, decorations.pathreplacing}
\begin{tikzpicture}[scale=0.65]
\draw[step=1, black, thick, opacity=0.1] (-7,-8) grid (8, 8);

\draw[thick, decorate,decoration={brace,amplitude=10}] (-2.95,-0.95) -- (-1.05, -0.95) node [midway, above, yshift=10] {$\gamma\, \Delta s$};
\draw[thick, decorate,decoration={brace,amplitude=10}] (-3.05,-2.95) -- (-3.05, -1.05) node [midway, anchor=east, xshift=-10] {$\gamma\,\Delta s$};

\draw[thick, decorate,decoration={brace,amplitude=10}] (1.95,2.05) -- (1.95,3.95) node [midway, anchor=east, xshift=-10] {$\gamma\,\Delta s$};
\draw[thick, decorate,decoration={brace,amplitude=10}] (2.05,4.05) -- (3.95,4.05) node [midway, anchor=south, yshift=10] {$\gamma\,\Delta s$};

\draw[thick, decorate,decoration={brace,amplitude=10}] (1.95,-3.95) -- (1.95,-1.05) node [midway, anchor=east, xshift=-10] {$\qty(\gamma+1)\,\Delta s$};
\draw[thick, decorate,decoration={brace,amplitude=10}] (3.95,-4.05) -- (2.05,-4.05) node [midway, anchor=north, yshift=-10] {$\gamma\,\Delta s$};

\fill[cbblue] (0,0) rectangle (1, 1);
\fill[cbblue, opacity=0.2] (-1,-1) rectangle (2, 2);

\foreach \x in {-1,...,2} {
  \foreach \y in {-1,...,2} {

    \draw[cbblue, fill=white, radius=0.1] (\x,\y) circle;

  }
}
\fill[cbblue, radius=0.06] node[anchor=north east] (0, 0) {$\mathbf{r}_0$} circle;

\fill[cbred] (-5,-5) rectangle (-4, -4);
\fill[cbred, opacity=0.2] (-6,-6) rectangle (-3, -3);

\foreach \x in {-6,-5,...,-3} {
  \foreach \y in {-6, -5, ..., -3} {
    \draw[cbred, fill=white, radius=0.1] (\x, \y) circle;
  }
}
\fill[cbred, radius=0.06] (-5, -5) node[anchor=north east, xshift=-3, yshift=-3] {$\mathbf{r}_a$} circle;

\fill[cbred] (5,5) rectangle (6, 6);
\fill[cbred, opacity=0.2] (4, 4) rectangle (7, 7);

\foreach \x in {4,...,7} {
  \foreach \y in {4,...,7} {
    \draw[cbred, fill=white, radius=0.1] (\x, \y) circle;
  }
}
\fill[cbred, radius=0.06] (5, 5) node[anchor=north east] {$\mathbf{r}_b$} circle;

\fill[cbgreen] (5,-6) rectangle (6, -5);
\fill[cbgreen, opacity=0.2] (4, -7) rectangle (7, -4);

\foreach \x in {4,...,7} {
  \foreach \y in {-7,...,-4} {
    \draw[cbgreen, fill=white, radius=0.1] (\x, \y) circle;
  }
}
\fill[cbgreen, radius=0.06] (5, -6) node[anchor=north east] {$\mathbf{r}_c$} circle;

\draw[thick, dashed] (-5.2,-5.2) rectangle (5.2, 5.2);

\end{tikzpicture}
  \caption{\label{fig:nearfield criterion}Illustration of the nearfield criterion for a third order expansion and $\gamma = 2$.
    The dashed line indicates the complete nearfield of the box associated with \textcolor{cbblue}{$\vb{r}_0$}---i.e.\ all boxes that have an expansion point within $\gamma \Delta s$ (infinity norm) of the expansion around \textcolor{cbblue}{$\vb{r}_0$}.
    Consequently, all of the $\vb{s}_\ell(\vb{r})$ within the central dark blue square have a pairwise interaction with the $\vb{s}_{\ell'}(\vb{r})$ inside the dashed box.
  }
\end{figure}
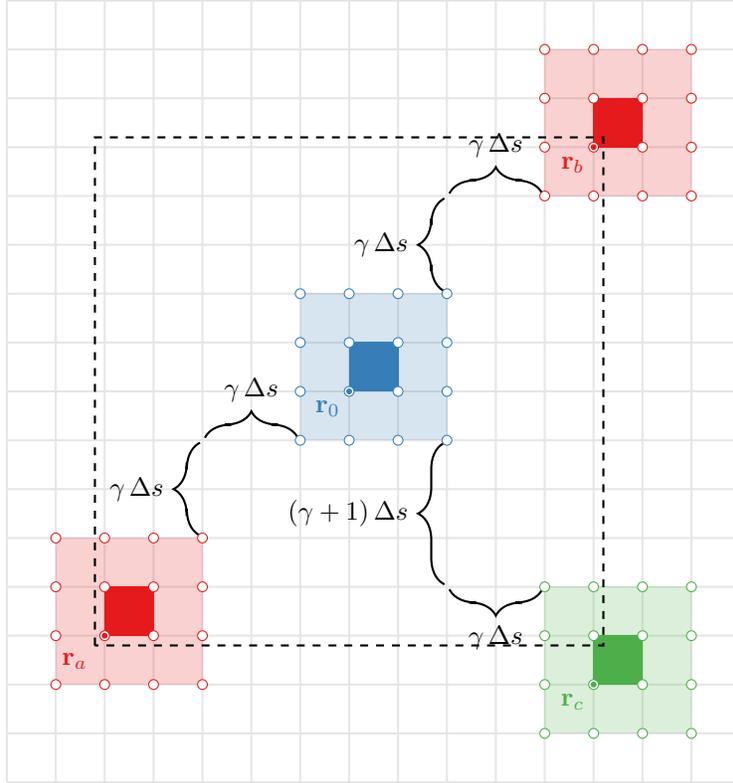

\subsubsection{\label{sec:expansion matrices}Auxiliary matrices}

The construction of both $\Lambda^\dagger$ and $\Lambda_\mathfrak{F}$ critically underpins the above process. These operators map quantum dot onto the uniform grid and back, though the operator $\Lambda_\mathfrak{F}$ differs slightly from $\Lambda^\dagger$ as it accounts for all the derivatives contained within $\mathfrak{F}$.
To start, we represent the primary $\vb{s}_\ell(\vb{r})$ basis functions as a weighted sum of $\delta$-functions on the surrounding gridpoints, thus $\vb{u}_a(\vb{r}) \propto \delta(\vb{r} - \vb{r}_a)$ and
\begin{equation}
  \psi_\ell(\vb{r}) \approx \sum_{\vb{u} \in C_\ell} \Lambda_{\ell\vb{u}}^\dagger \delta(\vb{r} - \vb{u}).
  \label{eq:grid linear combination}
\end{equation}
Here, $\psi_\ell(\vb{r}) \in \qty{\vb{s}_\ell(\vb{r})\cdot \vu{x}, \vb{s}_\ell(\vb{r}) \cdot \vu{y}, \vb{s}_\ell(\vb{r}) \cdot \vu{z}}$ and $C_\ell$ denotes the collection of grid points within the expansion region of $\vb{s}_\ell(\vb{r})$ (\cref{fig:aim terminology}).
For an expansion of order $M$, this sum contains $(M + 1)^3$ terms corresponding to the $(M + 1)^3$ grid points nearest to $\vb{s}_\ell(\vb{r})$. 
Consequently, the $\Lambda_{\ell \vb{u}}^\dagger$ matrices contain few nonzero elements and we have elected to use a moment-matching scheme to capture the $(M + 1)^3$ multipole moments of $\vb{s}_\ell(\vb{r})$ according to
\begin{equation}
  \sint (x - x_0)^{m_x}(y - y_0)^{m_y}(z - z_0)^{m_z} \qty[\psi_\ell(\vb{r}) - \sum_{\vb{u} \in c_\ell} \Lambda_{\ell\vb{u}}^\dagger \delta(\vb{r} - \vb{u})] \dd[3]{\vb{r}} = 0.
  \label{eq:moment matching}
\end{equation}
In this expression, $0 \leqslant m_x, m_y, m_z \leqslant M$ and $\vb{r}_0 \equiv x_0 \vu{x} + y_0 \vu{y} + z_0\vu{z}$ denotes the origin about which we compute the multipoles. To determine the $\Lambda_{\ell \vb{u}}^\dagger$, we  solve the least-squares system
\begin{equation}
  \sum_{\vb{u} \in C_\ell} W_{\vb{m}\vb{u}}\Lambda_{\ell\vb{u}}^\dagger = Q_{\ell\vb{m}}
  \label{eq:expansion matrix system}
\end{equation}
where
\begin{subequations}
  \begin{align}
    W_{\vb{m}\vb{u}} &= (u_x - x_0)^{m_x} (u_y - y_0)^{m_y} (u_z - z_0)^{m_z} \label{eq:w matrix} \\
    Q_{\ell \vb{m}} &= \int_{} \psi_\ell(\vb{r}) (x - x_0)^{m_x} (y - y_0)^{m_y} (z - z_0)^{m_z} \dd[3]{\vb{r}} \label{eq:q vector},
  \end{align}
\end{subequations}
$\vb{u} \in C_\ell$, and $\vb{m}$ denotes the multi-index $\vb{m} = \qty{m_x, m_y, m_z}$.
With an infinite precision calculation, the choice of $\vb{r}_0 = x_0 \vu{x} + y_0 \vu{y} + z_0 \vu{z}$ merely defines an origin for the polynomial expansion system.
To minimize numerical issues, we choose $\vb{r}_0$ at the center of $\vb{s}_\ell(\vb{r})$.

As $\Lambda^\dagger$ arises purely as a function of space with no time component, differentiating \cref{eq:moment matching} with respect to $x$, $y$, or $z$ amounts to differentiating the polynomial that interpolates $\mathfrak{F}\qty{\vb{P}(\vb{r}, t)}$ between grid points and thus can recover spatial derivatives occurring in $\mathfrak{F}$. As a result, differentiating \cref{eq:moment matching} removes the high-order moments in \cref{eq:q vector}.

\subsection{Convergence analysis}

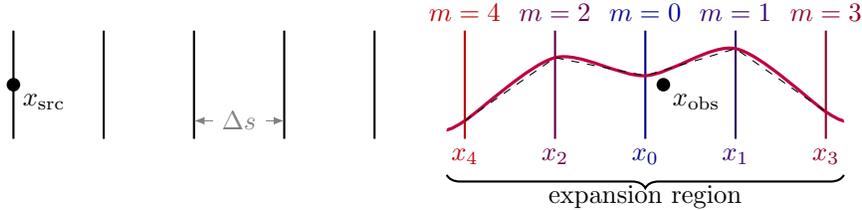
\begin{figure}
    \centering
    \usetikzlibrary{decorations.markings}
\usetikzlibrary{decorations.pathreplacing}
\tikzset{
    mark position/.style args={#1(#2)}{
        postaction={
            decorate,
            decoration={
                markings,
                mark=at position #1 with \coordinate (#2);
            }
        }
    }
}

\begin{tikzpicture}[scale=1.2, >=latex]

\foreach \i in {0, ..., 4}
{
  \draw[thick] (\i, -0.6) -- (\i, 0.6);
}

\fill[] (0,0) circle (0.075) node [anchor=north west] {$x_\text{src}$};
\fill[] (7.2,0) circle (0.075) node [anchor=north west] {$x_\text{obs}$};

\draw[thick, blue!75!black!100!red!75!black] (7,-0.6) node[below]{$x_0$} -- (7, 0.6) node[above]{$m = 0$};
\draw[thick, blue!75!black!75!red!75!black] (8,-0.6) node[below]{$x_1$} -- (8, 0.6) node[above]{$m = 1$};
\draw[thick, blue!75!black!50!red!75!black] (6,-0.6) node[below]{$x_2$} -- (6, 0.6) node[above]{$m = 2$};
\draw[thick, blue!75!black!25!red!75!black] (9,-0.6) node[below]{$x_3$} -- (9, 0.6) node[above]{$m = 3$};
\draw[thick, blue!75!black!0!red!75!black] (5,-0.6) node[below]{$x_4$} -- (5, 0.6) node[above]{$m = 4$};

\draw[very thick, purple] plot [smooth] coordinates {(4.8, -0.5) (5,-0.4) (6, 0.3) (7, 0.1) (8, 0.4) (9, -0.3) (9.2, -0.4)};
\draw[dashed] plot [] coordinates {(5,-0.4) (6, 0.3) (7, 0.1) (8, 0.4) (9, -0.3)};

\draw[<->, gray] (2,-0.4) -- (3,-0.4) node[midway, fill=white] {$\Delta s$};
\draw[thick, decorate,decoration={brace,amplitude=5}] (9.2,-1)  -- (4.8,-1)  node [black,midway,below=0.2] {expansion region};

\end{tikzpicture}
    \caption{\label{fig:1d moments} Polynomial interpolation of $g(x - x_\text{src})$ near $x_\text{obs}$.
    Here, the green curve represents the actual $g(x - x_\text{src})$ and the dashed black line its approximation.
    Evaluating the $m^\text{th}$-order approximation requires samples of the signal at $m + 1$ grid points surrounding $x_\text{obs}$.}
\end{figure}

Next, we present a succinct analysis of convergence. While such analyses arise in different contexts \cite{Bleszynski1996}, the analysis herein approaches it from an interpolation perspective and enables one to obtain an error bound on the overall operator.  With no loss of generality, consider two point particles located at $x_\text{src}$ and $x_\text{obs}$.
A time-independent Green's function, $g(x_\text{obs} - x_\text{src})$, describes the interaction between the two particles and we wish to construct a polynomial approximation of $g(x - x_\text{src})$ for $x$ in the vicinity of $x_\text{obs}$ as in \cref{fig:1d moments}. 

To construct an interpolation polynomial over the expansion region of order $M$, we define a polynomial co\"ordinate $x_p= \qty(x - x_0)/\Delta s$ in units of $\Delta s$ such that $x_p^\text{min} \leqslant x_p \leqslant x_p^\text{min} + M$ where $x_p^\text{min} \equiv -\floor*{M/2}$. 
Consequently, the expansion points about $x_\text{obs}$ correspond to $x_p \in \qty{-\floor{M/2}, -\floor{M/2} + 1, -\floor{M/2} + 2, \ldots}$ with the \nth{0} order expansion point, $x_0$, equivalent to $x_p = 0$.
Such a co\"ordinate system defines the Vandermonde's linear equation $\sum_{j}V_{ij} w_j = g_i$ for the weights of an interpolating polynomial where
\begin{subequations}
  \begin{align}
    V_{ij} &= (x_p^\text{min} + i)^j \\
    g_i &= g\qty((x_0 - x_\text{src}) + (x_p^\text{min} + i)\, \Delta s)
  \end{align}
\end{subequations}
and $0 \leqslant i, j \leqslant M$.
Approximating $g(x - x_\text{src})$ at $x_\text{obs}$ then becomes a matter of evaluating this polynomial at $x_p = \qty(x_\text{obs} - x_0)/\Delta s$, i.e.
\begin{equation}
  g(x_\text{obs} - x_\text{src}) = g\qty((x_0 - x_\text{src}) + \qty(\frac{x_\text{obs} - x_0}{\Delta s}) \Delta s ) \approx \sum_{i = 0}^{M} w_i \qty(\frac{x_\text{obs} - x_0}{\Delta s})^i.
\end{equation}
Accordingly, the polynomial approximation to $g(x_\text{obs} - x_\text{src})$ contains terms of order $\mathcal{O}(\Delta s^{-M})$ and we can expect the approximation error to scale as $\mathcal{O}(\Delta s^{-(M + 1)})$.
This also motivates using the approximation to calculate interactions involving differential operators; applying an $n$th-order derivative reduces the polynomial order by $n$, thus the error scales like $\mathcal{O}(\Delta s^{-(M + 1) + n})$. The preceding analysis generalizes to three dimensions. 

\section{Numerical results}\label{sec:results}

Next, we present a number of result using the methodologies developed in this paper. We seek to demonstrate controllable accuracy of the proposed scheme, the cost complexity in both computation time and memory, and finally some exemplar simulations of \qd{} systems. 

\subsection{Accuracy}
\label{sec:numerical accuracy}

To start, we examine error incurred in our approach in evaluating the space time convolution in \eqref{eq:rotatedRadFld}. To isolate the errors incurred, our experiment proceeds as follows. We set up two domains with sufficient separation such that the interactions between these occur only via AIM. Each domain contains 64 randomly distributed \qds{}, we prescribe the temporal variation of the polarization of each \qd{}, and we measure the total radiated field at each \qd{}. Finally, we fix the temporal interpolation basis order at 3 and the polarization of each \qd{} varies as
\begin{equation}\label{eq:errortest gaussian}
    P(t) = e^{-\frac{(t - t_0)^2}{2 \sigma^2}}
\end{equation}

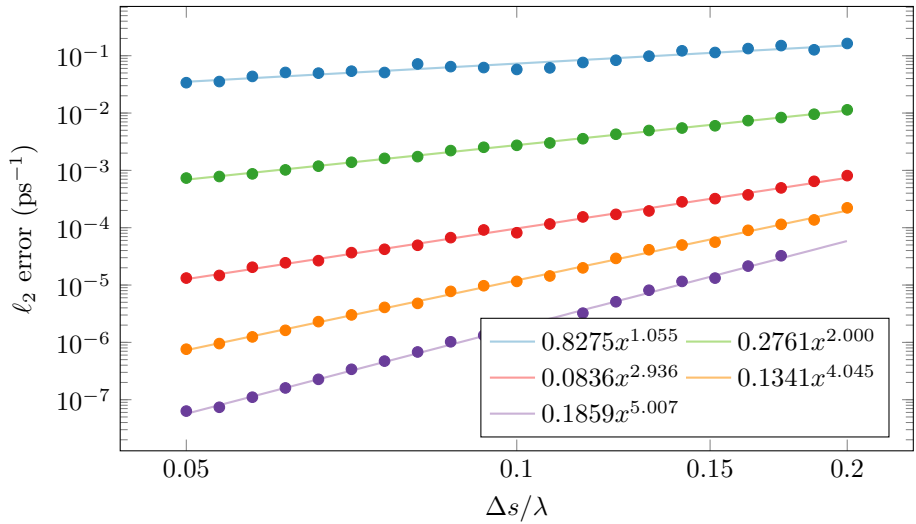
\begin{figure}
    \centering
    \pgfplotsset{cycle list/Paired-10}
\begin{tikzpicture}
  \begin{loglogaxis}[
    cycle multi list={Paired-10},
    xlabel=$\Delta s/\lambda$,
    ylabel=$\ell_2$ error (\si{\per\pico\second}),
    legend pos=south east,
    legend columns=2,
    legend entries={$0.8275x^{1.055}$,,$0.2761x^{2.000}$,,$0.0836x^{2.936}$,,$0.1341x^{4.045}$,,$0.1859x^{5.007}$},
    xtick={0.05, 0.1, 0.15, 0.2},
    xticklabels={0.05, 0.1, 0.15, 0.2},
    width=\linewidth,
    height=0.618034\linewidth
  ]
    \addplot+[thick, domain = 0.05:0.2, samples = 128] {0.827533*x^1.05491};
    \addplot+[only marks] table[x index = 0, y index = 1] {static_error.dat};

    \addplot+[thick, domain = 0.05:0.2, samples = 128] {0.276082*x^2.00038};
    \addplot+[only marks] table[x index = 0, y index = 2] {static_error.dat};

    \addplot+[thick, domain = 0.05:0.2, samples = 128] {0.083601*x^2.93566};
    \addplot+[only marks] table[x index = 0, y index = 3] {static_error.dat};

    \addplot+[thick, domain = 0.05:0.2, samples = 128] {0.134061*x^4.04545};
    \addplot+[only marks] table[x index = 0, y index = 4] {static_error.dat};

    \addplot+[thick, domain = 0.05:0.2, samples = 128] {0.185903*x^5.00736};
    \addplot+[only marks] table[x index = 0, y index = 5] {static_error.dat};
  \end{loglogaxis}
\end{tikzpicture}
    \caption{$l_2$ error of the Rabi frequency magnitude $\abs{\chi}$ with respect to grid spacing for expansion orders 2 through 6, using source and observer boxes of volume $\lambda^3$ separated by $\Delta r = 2\lambda (\vu{x} + \vu{y} + \vu{z})$, each containing 64 randomly generated \qds{}. For an expansion order $M$ one expects the overall error to scale as $O(\Delta s^{M-1})$, consistent with the results above.}
    \label{fig:static error test}
\end{figure}

The simulation runs for 1024 timesteps of size $\Delta t = \SI{0.1}{\pico\second}$, the width of the Gaussian $\sigma = 1024 \, \Delta t / 12$ and its center $t_0 = 1024 \, \Delta t / 2$. This approach admits a readily available analytic solution via \cref{eq:integral operator} which we measure against the AIM solution. For this, we calculate the $\ell_2$ norm differences between the two solutions as a function of AIM grid size for different expansion orders to validate the error behavior described in \cref{sec:AIM}. \Cref{fig:static error test} gives geometric parameters and results; as shown by the figure, we observe excellent convergence. 

\begin{figure}
    \centering
        \includegraphics[width=.49\textwidth]{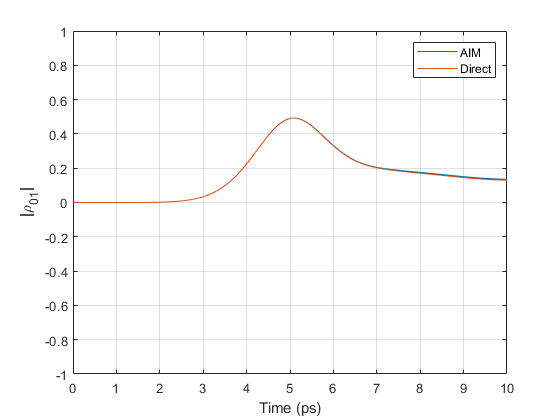}\includegraphics[width=.49\textwidth]{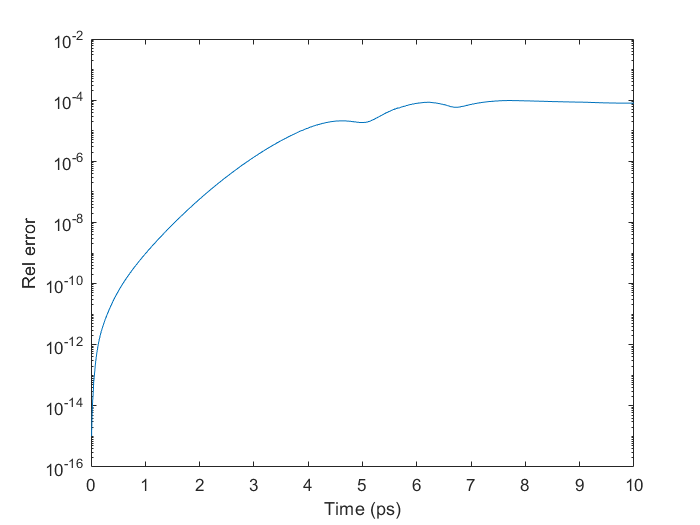}
   \caption{(Left) $|\tilde{\rho}_{01}|$ for a system of 128 \qds{} in a cube of length $\lambda/10$ (chosen to promote observable coupling effects through very close \qds{}), computed using the direct algorithm and AIM. (Right) Relative error of AIM algorithm against the direct algorithm for this simulation.}
    \label{fig:dynamic error test}
\end{figure}

Next, we examine errors incurred when conducting a similar experiment in the rotating frame. All \qds{} begin in the ground state $(\rho_{00}, \rho_{01})|_{t=0} = (1,0)$, and their density matrix elements evolve according to \cref{eq:rotating liouville}. The dipole moment of each \qd{} aligns with the laser field, given by
\begin{equation}
    \tilde{\vb{E}}(\vb{r},t)=  \tilde{E}_0 \: \vu{x} \: e^{-\frac{(\vb{k} \cdot \vb{r}-\omega(t-t_0))^2}{2 \sigma^2}}.
\end{equation}
We use a fifth order expansion with AIM spacing $\Delta s = 5 \times 10^{-3} \lambda$, and 1000~timesteps of size $\Delta t = \SI{0.01}{\pico\second}$ (\cref{tab:params} gives additional simulation parameters.) As before, we compare results from AIM (\cref{fig:dynamic error test}) to those obtained using the direct method, as it permits us to normalize against the error in using temporal basis sets. 

\begin{table}[]
    \centering
    \begin{tabular}{lll}
        Quantity & Symbol & Value \\
        \hline
        Speed of light & $c$ & \SI{300}{\micro\meter\per\pico\second}  \\
        Transition frequency  & $\omega_0$ & \SI{1500}{\milli\eV \per \hbar} \\
        Transition dipole moment (magnitude) & $|\vec{d}|$ & $10 \: e a_0$   \\
        Decoherence times & $T_1$, $T_2$ & \SI{10}{\pico\second}, \SI{20}{\pico\second} \\
        Laser frequency & $\omega$ & \SI{1500}{\milli\eV \per \hbar} \\
        Laser wavevector & $|\vb{k}|$ & \SI{7.6016}{\per \micro\meter} \\
        Laser wavelength & $\lambda$ & \SI{827}{\nano\meter} \\
        Laser peak shift & $t_0$ & \SI{5}{\pico\second} \\
        Pulse width & $\sigma/\omega$ & \SI{1}{\pico\second} \\
        Pulse area & - & $\pi$ \\
    \end{tabular}
    \caption{Dynamic simulation parameters; $e$ and $a_0$ denote the elementary charge and Bohr radius. The decoherence times here, while shorter than those typical of optical resonance experiments, afford a shorter computational time but preserve dynamical emission phenomena.}
    \label{tab:params}
\end{table}

\subsection{Cost of evaluation of higher order spatial derivatives}

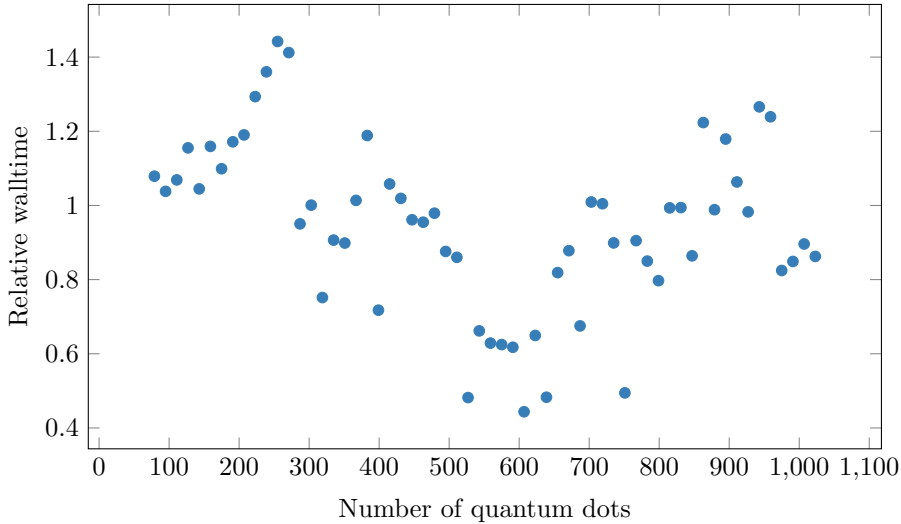
\begin{figure}
    \centering
    \begin{tikzpicture}
  \begin{axis}[
      xlabel = {Number of \qds},
      ylabel = {Relative walltime},
      width=\linewidth,
      height=0.618034\linewidth
    ]
    \addplot+[only marks, color = cbblue] table {kernel_time_comparison.dat};
  \end{axis}
\end{tikzpicture}
    \caption{\label{fig:kernel time comparison}Simulation time of $g(\vb{r}, t) \ast \tilde{\vb{P}}(\vb{r}, t)$ relative to $\tilde{\mathfrak{F}}\{\tilde{\vb{P}}(\vb{r}, t)\}$ for various system sizes.}
\end{figure}

Figure \ref{fig:kernel time comparison} shows walltime results for the calculation of $g(\vb{r}, t) \ast \tilde{\vb{P}}(\vb{r}, t)$ relative to $\tilde{\mathfrak{F}}\{\tilde{\vb{P}}(\vb{r}, t)\}$---\ie\ a ``simple'' scalar propagator relative to a complex one involving dyadics and derivatives---for various system sizes.
Each experiment uses the same configuration of sources (arranged linearly at consistent density) and system parameters and we time only the timestepping procedure assuming pre-filled matrices. 
We attribute the correlated variation in \cref{fig:kernel time comparison} to AIM---the efficiency of the grid-based acceleration scheme accutely depends on the geometry/density of sources---though we note both propagators appear to take roughly the same amount of computational effort to evaluate.
This indicates that our modified TD-AIM formulation can accommodate \emph{any} propagation kernel involving arbitrary spatiotemporal derivatives with little-to-no additional computational overhead.

\subsection{Complexity}

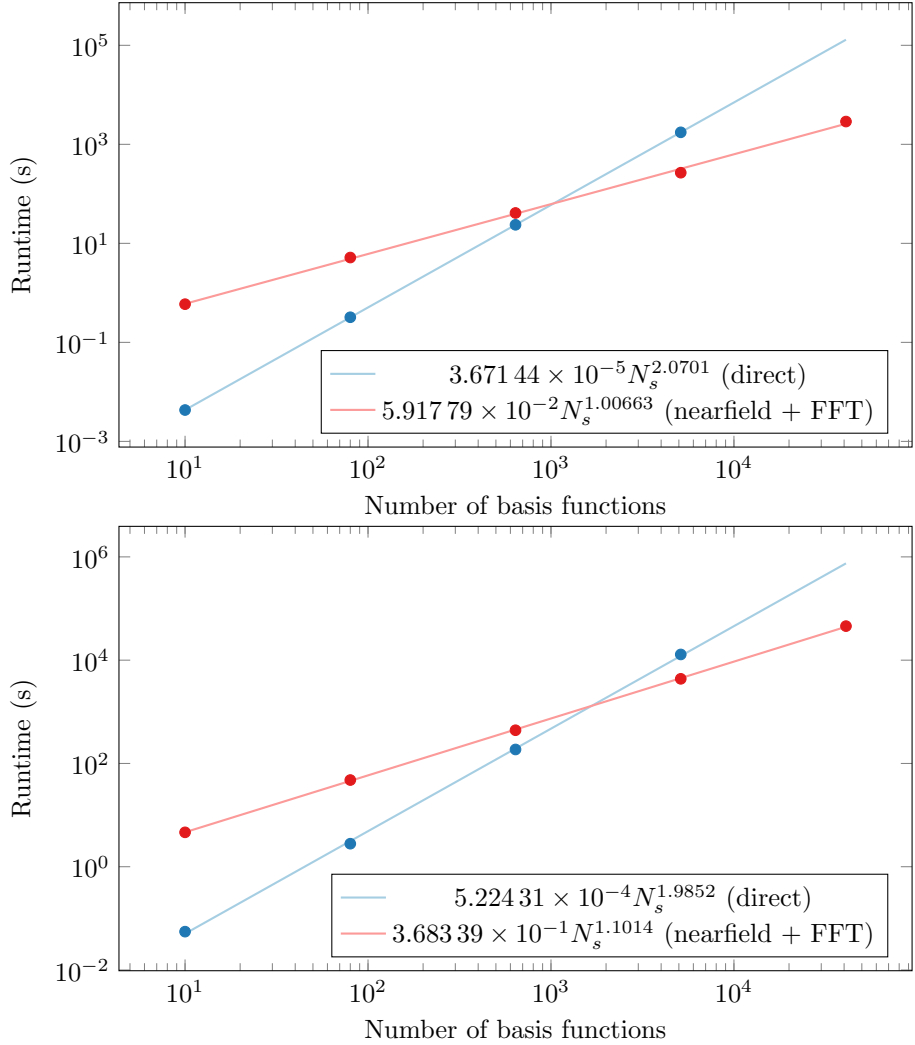
\begin{figure}
    \centering
    \pgfplotsset{cycle list/Paired-10}
\begin{tikzpicture}
  \begin{loglogaxis}[
    cycle multi list={Paired-10},
    xlabel=Number of basis functions,
    ylabel=Runtime (\si{\second}),
    legend pos=south east,
    legend entries={$\num{3.67144E-5}N_s^{2.0701}$ (direct),,$\num{5.91779E-2}N_s^{1.00663}$ (nearfield + FFT)},
    width=\linewidth,
    height=0.618034\linewidth
  ]
    \addplot+[thick, domain = 10:40960, samples = 32] {0.0000367144*x^2.07013};
    \addplot+[only marks] table[x index = 0, y index = 1] {static_runtime.dat};

    \pgfplotsset{cycle list shift=2}

    \addplot+[thick, domain = 10:40960, samples = 32] {0.0591779*x^1.00663};
    \addplot+[only marks] table[x index = 0, y index = 2] {static_runtime.dat};
  \end{loglogaxis}
\end{tikzpicture}

\begin{tikzpicture}
  \begin{loglogaxis}[
    cycle multi list={Paired-10},
    xlabel=Number of basis functions,
    ylabel=Runtime (\si{\second}),
    legend pos=south east,
    legend entries={$\num{5.22431E-4}N_s^{1.9852}$ (direct),,$\num{3.68339E-1}N_s^{1.1014}$ (nearfield + FFT)},
    width=\linewidth,
    height=0.618034\linewidth
  ]
    \addplot+[thick, domain = 10:40960, samples = 32] {0.000522431*x^1.98518};
    \addplot+[only marks] table[x index = 0, y index = 1] {dynamic_runtime.dat};

    \pgfplotsset{cycle list shift=2}

    \addplot+[thick, domain = 10:40960, samples = 32] {0.368339*x^1.10143};
    \addplot+[only marks] table[x index = 0, y index = 2] {dynamic_runtime.dat};
  \end{loglogaxis}
\end{tikzpicture}
    \caption{FFT runtime (excluding setup time) using a third-order expansion. (Top) 1024 timesteps with $\Delta s = \lambda / 400$ and prescribed polarizations in the fixed frame. (Bottom) 1000 timesteps with $\Delta s = \lambda / 10$ and Liouville-dynamics polarization in the rotating frame. Both cases have a quasi-quadratic scaling in the direct calculation, whereas the FFT-accelerated calculation performs slightly worse than linear.}
    \label{fig:perfAIMstatic}
\end{figure}

Next, we present a set of experiments that demonstrate the $O(N_s \log(N_s))$ complexity scaling of AIM. For this, we perform simulations in both the fixed frame with prescribed polarizations, and the rotating frame with full Liouville equation dynamics. To ensure proper examination of computational complexity, we start with a box of side length $6 \, \Delta s$ (chosen to minimize the number of nearfield pairs), and filled with \qds{} at random locations. We obtain each successive value of $N_s$ by doubling the sidelength and in effect, increasing the number of \qds{} by a factor of eight. We use a third order approximation with AIM spacings $\Delta s = \lambda / 400$ and $\Delta s = \lambda / 10$ for the fixed and rotating wave cases, respectively. Timesteps mirror those used in \cref{sec:numerical accuracy}. \Cref{fig:perfAIMstatic} gives runtimes for both cases, demonstrating that the two FFT-accelerated simulations outpace their direct counterparts near $N_s = 1000$ and $N_s = 2000$, respectively.

\begin{figure}
    \centering
    \includegraphics[width=\linewidth]{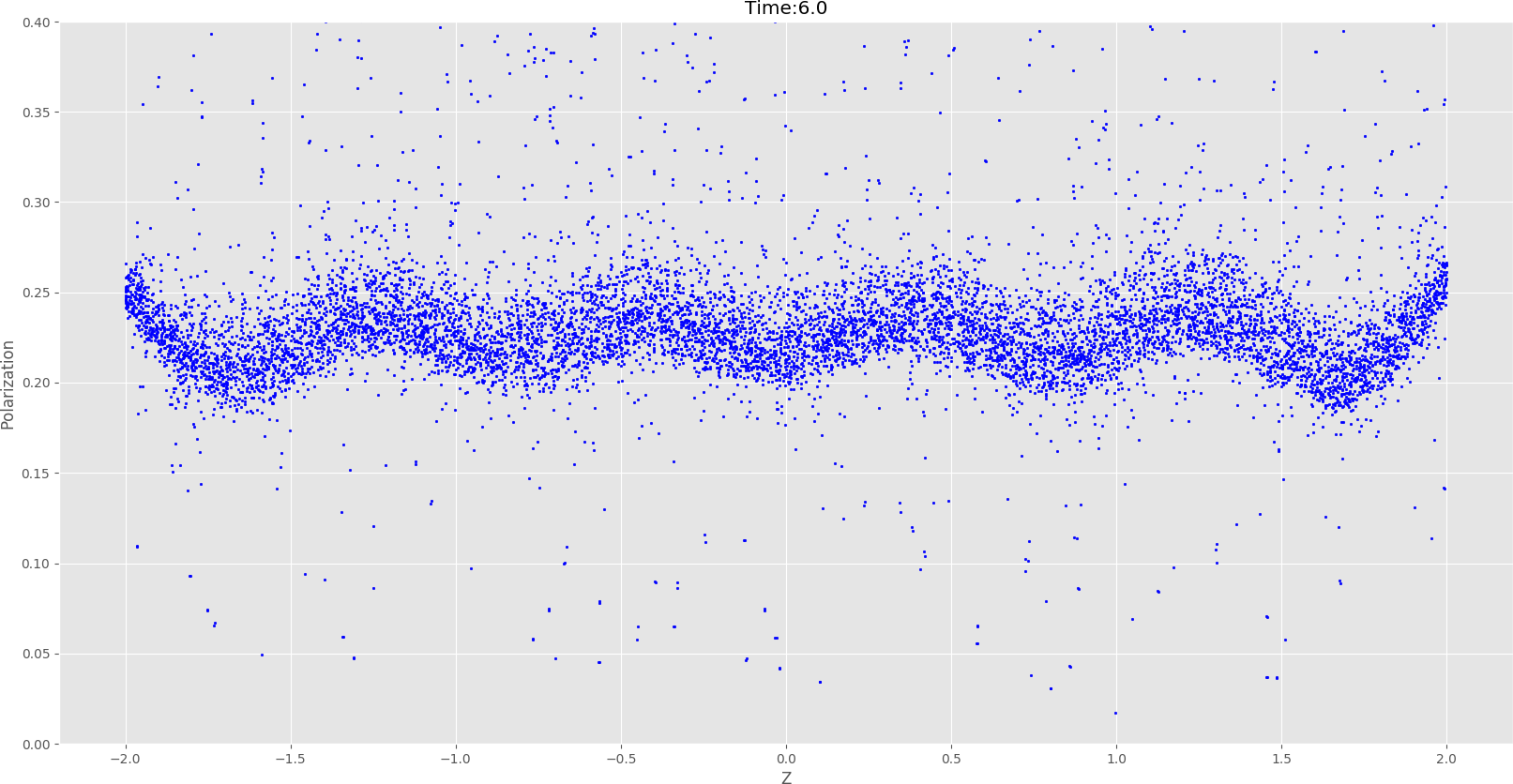}
    \caption{$\vu{z}$-distribution of polarization $|\tilde{\rho}_{01}|$ for a $\num{10000}$-dot cylindrical simulation, replicating the parameters in \cite{glosser2017collective}. The AIM calculation recovers the oscillatory long-range pattern that we obtained using a direct calculation \cite{glosser2017collective}.}
    \label{fig:10kcylinder}
\end{figure}

\subsection{Large scale physical simulations}
\label{subs:tow}
 The largest system simulated in \cite{glosser2017collective} without TD-AIM  consists of \num{10000} \qds{} randomly distributed in a cylinder of radius \SI{0.2}{\micro\meter} and length \SI{4}{\micro\meter}. \Cref{fig:10kcylinder} shows an equivalent simulation with TD-AIM that reproduces features arising from \qd{} interactions; we conduct a very similar experiment here. The figure shows the polarization of each \qd{} in the cylinder as a function of their $\vb{z}$-coordinate (the axis of the cylinder), under the effect of a resonant $\pi$ pulse. Each of the \qds{} has an identical (fixed) dipole moment (see Ref.\cite{glosser2017collective} for the details of the simulation parameters). Note how the secondary radiation produces random shifts in the polarization due to short-range effects in the local neighborhood of each \qd. In addition, the simulation shows an oscillation of the polarization due to long-range collective effects. This oscillation reflects the role of boundary conditions in the confinement of the macroscopic electric field in the system.

\begin{figure}
    \centering
    \includegraphics[width=\linewidth]{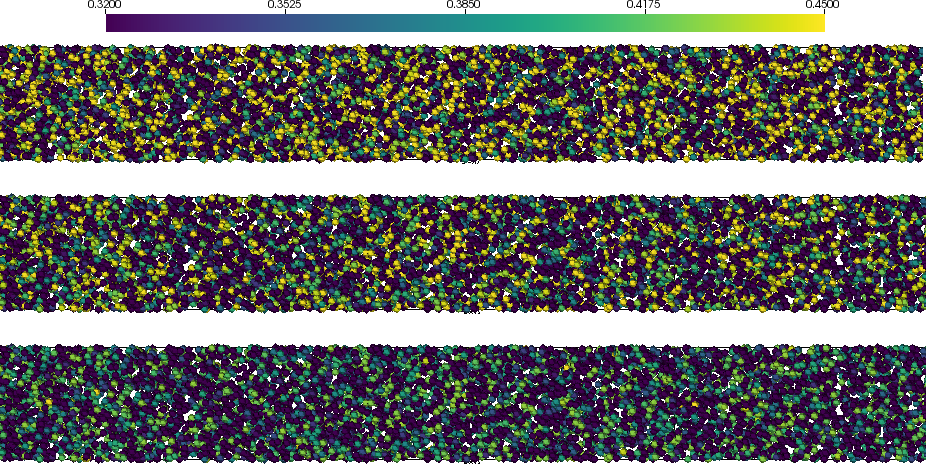}
    \caption{Coloration of $\abs{\tilde{\rho}_{01}}$ as an indicator of $|\tilde{\textbf{P}}|$ at $t_1 = \SI{2.0}{\pico\second}$ (top), $t_2 = \SI{3.0}{\pico\second}$ (middle), $t_3 = \SI{4.0}{\pico\second}$ (bottom) relative to the peak of a \SI{1}{\pico\second}-wide pulse, for a system of \num{100000} \qds{}. }
    \label{fig:100kpolcolor}
\end{figure}

\begin{figure}
    \centering
    \includegraphics[width=\textwidth]{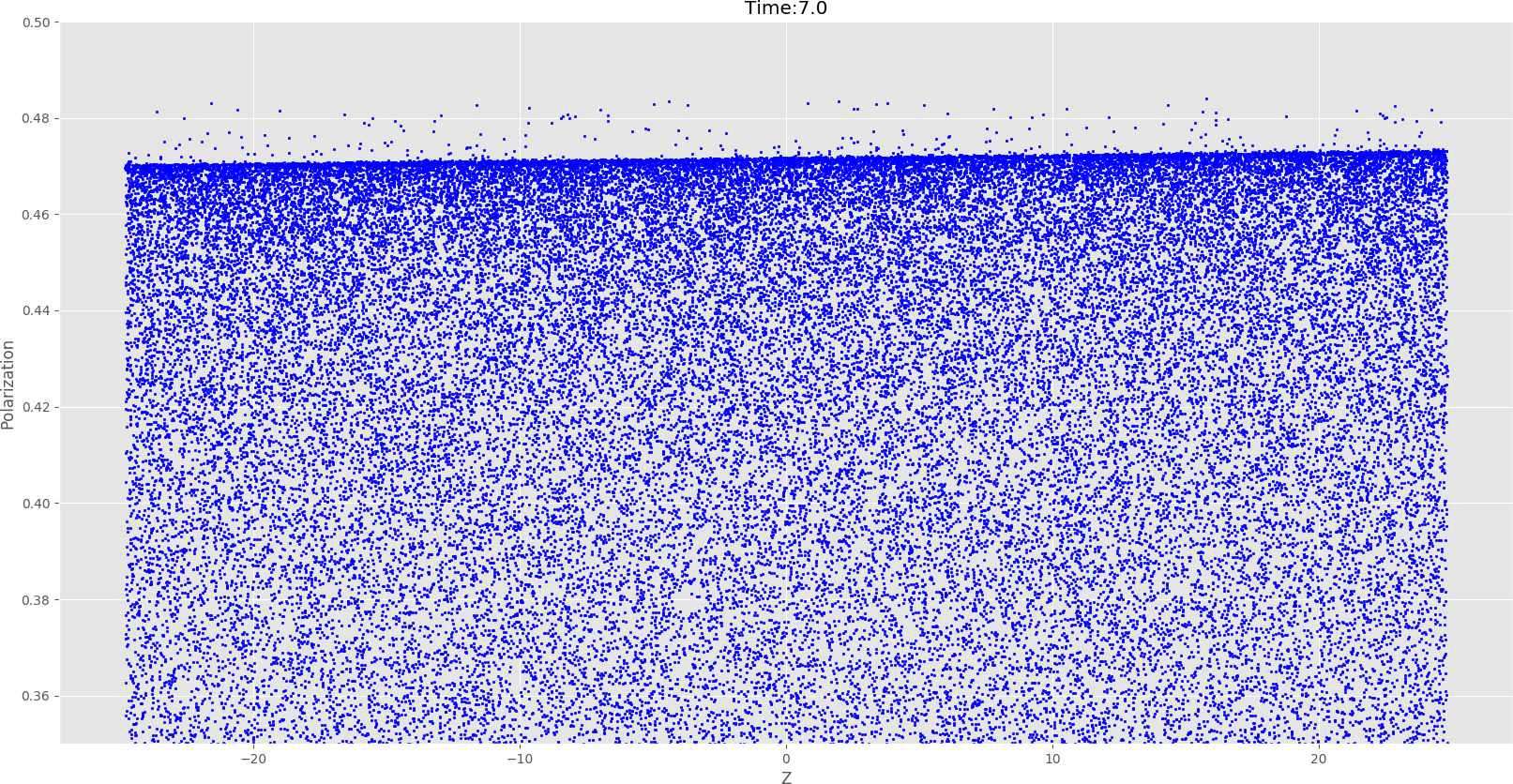}
    \includegraphics[width=\textwidth]{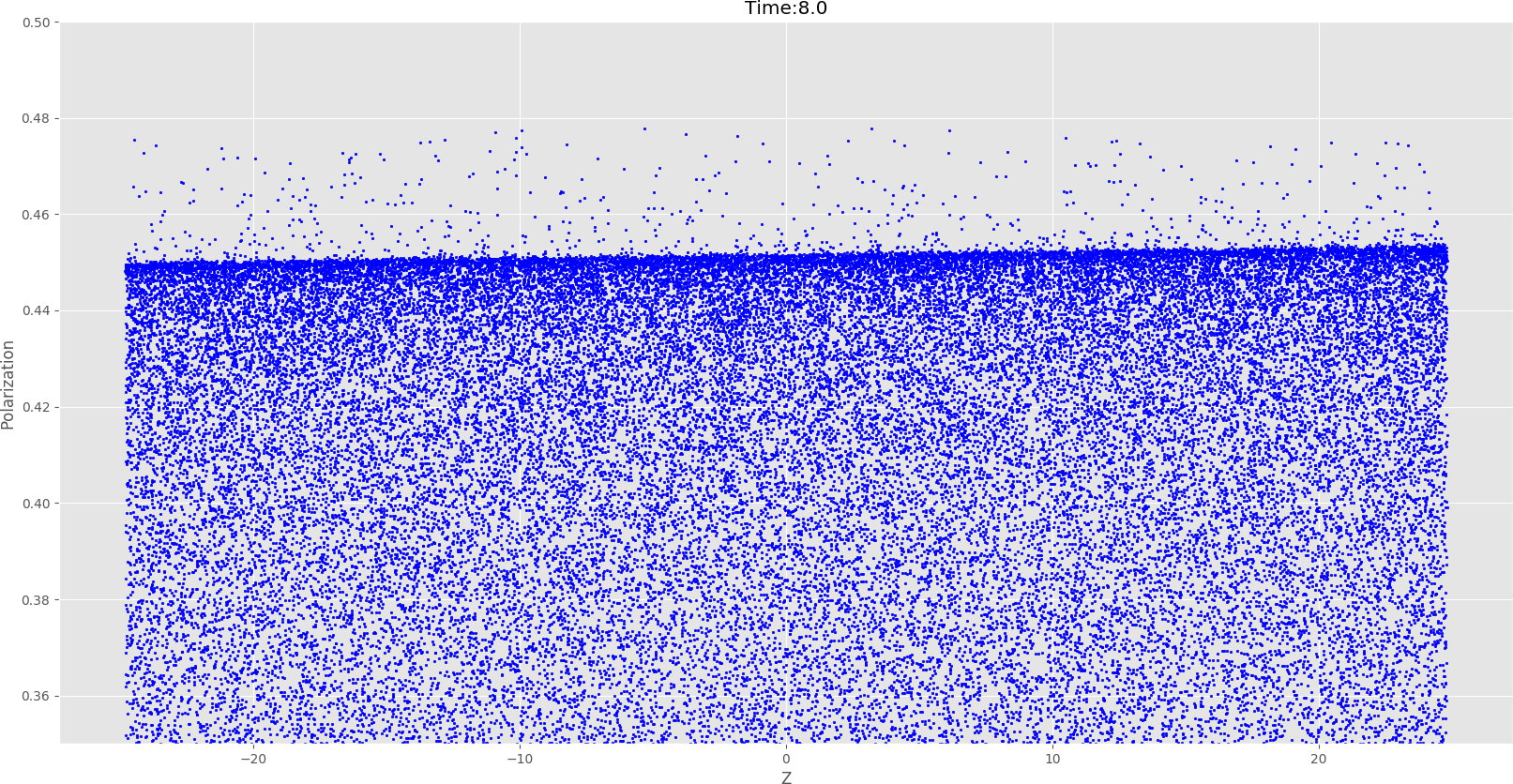}
    \includegraphics[width=\textwidth]{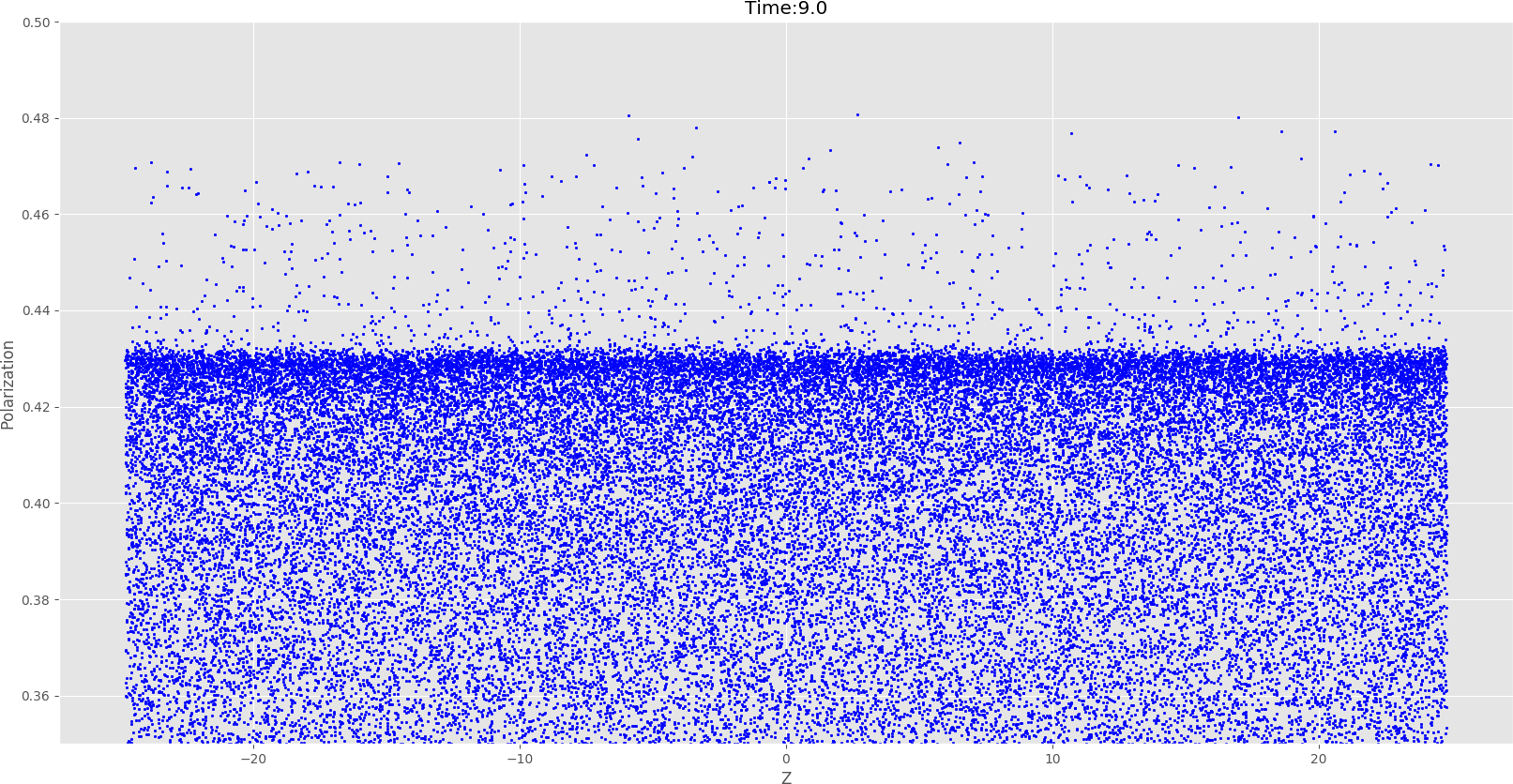}

    \caption{Scatterplots of $|\tilde{\rho}_{01}|$ corresponding to the color plots of \cref{fig:100kpolcolor}. There exists a single preferred polarization, represented by the linear region of greatest density, arising from \qds{} whose transition dipole moments (anti-)align with the laser field. Radiative coupling produces polarizations that exceeding this value.}
    \label{fig:100kpolscatter}
\end{figure}

The algorithm introduced in this paper facilitates simulations of much larger systems. In \cref{fig:100kpolcolor,fig:100kpolscatter} we examine the response of a system of \num{100000} \qds{}---randomly distributed throughout a cuboid---to an applied laser pulse traveling along $\vu{z}$. The transition dipole moment of each \qd{} has a fixed magnitude but random orientation. \Cref{tab:params,tab:100kparams} list simulation parameters.
\begin{table}[]
    \centering
    \begin{tabular}{lll}
        Quantity & Symbol & Value \\
        \hline
        Simulation timestep & $\Delta t$ & $0.02$ ps \\
        AIM spacing & $\Delta s$ & $0.040 \lambda = 33.06$ nm \\
        Transverse domain length & - & $16\: \Delta s = 529$ nm \\
        Longitudinal domain length & - & $1500 \: \Delta s = 49.59 \:\mu$m
    \end{tabular}
    \caption{AIM parameters for the simulation of Section \ref{subs:tow}}
    \label{tab:100kparams}
\end{table}

Fig~\ref{fig:100kpolcolor} displays a color map of $\abs{\tilde{\rho}_{01}}$ as an indicator of the polarization $|\tilde{\textbf{P}}|$ of each \qd{} at different timesteps after the pulse peak. The figure shows only \qds{} located in a central segment of about \SI{4}{\micro\meter} of the entire cuboid. The random orientation of the dipole moments creates a variation in the amplitude of the polarization with \qds{} whose dipole moments (anti-)align with the laser field having greatest amplitude. In addition, despite each \qd{} resonantly coupling to the pulse, inhomogeneity arises due to the inter-dot coupling. These simulation can resolve inhomogeneities at the microscopic level, taking into account the orientation of the transition dipole moment of each \qd{}, as well as the effect of local secondary fields.

To visualize long-range effects, \cref{fig:100kpolscatter} shows $|\tilde{\rho}_{01}|$ as a function of the $\vb{z}$ coordinate of each \qd{}, corresponding to the color plots of \cref{fig:100kpolcolor}. Here we show the entire cuboid having sides of \SI{20}{\micro\meter}. In contrast to the results of \cref{fig:10kcylinder}, we do not observe the oscillatory behavior due to confinement since the length of the system far exceeds the radiation wavelength. Moreover, we observe a dispersion of the polarization due to the random orientation of the transition dipoles. Since the strength of the coupling scales with $\vb{E} \cdot \vb{d} = \cos(\theta)$, the distribution peaks at the value of $|\tilde{\rho}_{01}|$ when $\theta=0$ or $\theta=\pi$, with a tail corresponding to all the intermediate values. Only a few \qds{}, for which the secondary fields constructively interfere, have a polarization larger than the peak value. Finally, note how the value of the peak polarization slightly increases from left to right due to pulse propagation.

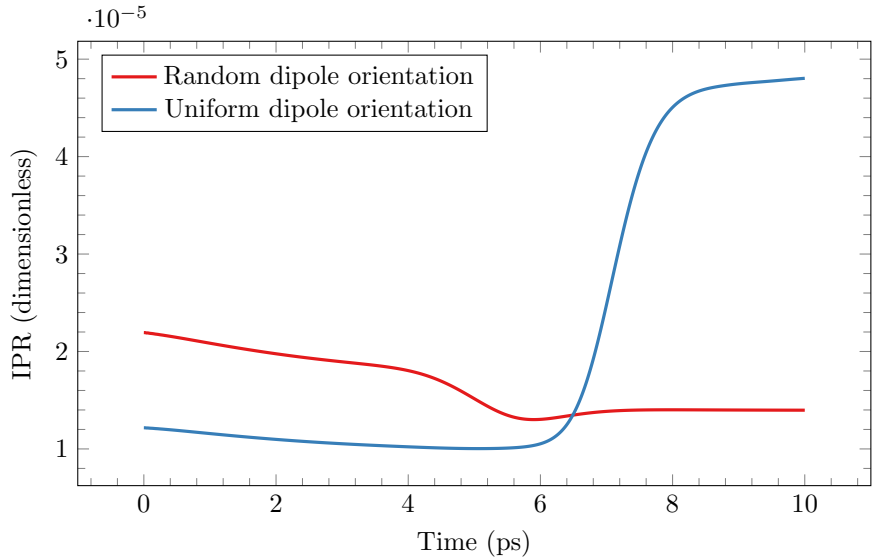
\begin{figure}
    \centering
    \pgfplotsset{cycle list/Set1-3}
\begin{tikzpicture}
  \begin{axis}[
    cycle multi list={Set1-3},
    xlabel=Time (\si{\pico\second}),
    ylabel=IPR (dimensionless),
    legend pos=north west,
    minor tick num = 4,
    width=\linewidth,
    height=0.618034\linewidth
  ]
    \addplot+[very thick, no marks] table[x index = 0, y index = 1] {ipr.dat};
    \addlegendentry{Random dipole orientation}

    \addplot+[very thick, no marks] table[x index = 0, y index = 2] {ipr.dat};
    \addlegendentry{Uniform dipole orientation}
  \end{axis}
\end{tikzpicture}
    \caption{Inverse Participation Ratio (IPR) for the system of \cref{fig:100kpolcolor} (red) and a similar system of \num{100000} \qds{} with uniform dipole orientations (blue).}
    \label{fig:IPR}
\end{figure}

Furthermore, we calculate the inverse participation ratio (IPR) of the dot polarization:
\begin{equation}
    \text{IPR}(t) = \frac{\sum_l |\tilde{\rho}_{01}(t)|^4}{(\sum_l |\tilde{\rho}_{01}(t)|^2)^2}
\end{equation}
with results shown in \cref{fig:IPR}. This quantity ranges from $1/N$ to $1$, and collectively measures the spatial localization of the polarization, with $1/N$ corresponding to a completely delocalized spatial distribution, and $1$ to the case of the polarization completely localised on as single site.  For comparison, we also include an IPR plot for the case of uniform (pulse-aligned) dipoles. In the uniform case, all \qds{} participate equally until the onset of the pulse peak, whereupon inter-dot coupling leads certain \qds{} to retain their polarization longer than neighbors. This contrasts the non-uniform case, which exhibits localization of polarization to \qds{} that align with the laser pulse.

\section{Conclusions}\label{sec:conclusions}
Here we have presented novel variations to TD-AIM that enables analysis of large ensembles of \qds.
We discuss Numerous features of the approach, including accuracy, convergence, and complexity.
The latter for prescribed and fixed polarization, as well as when the polarization evolves.
We validate the approach against ``direct'' simulations that use no acceleration techniques.
Finally, we use the approach simulate a system with \num{100000} dots.
We observe identical results identical to direct solutions, thus these techniques can reliably simulate much larger systems.
The next phase of our research focuses on additional functionality in both the physics as well as the computational infrastructure.

\section*{Acknowledgments}
We greatfully acknowledge the HPCC facility at Michigan State University for their support of this work.
We also acknowledge support from the National Science Foundation under grants NSF ECCS-1408115 and OAC-1835267.

\section*{References}
\bibliography{refs,random_lasing_reflist}

\end{document}